\documentclass[authoryear]{cas-dc}

\usepackage[numbers]{natbib}
\usepackage{hyperref}

\usepackage{amssymb}
\usepackage{lipsum}
\usepackage{ragged2e}
\usepackage{orcidlink}

\hypersetup{
    colorlinks=true,
    citecolor=blue,  
    linkcolor=blue,  
    urlcolor=blue    
}

\def\tsc#1{\csdef{#1}{\textsc{\lowercase{#1}}\xspace}}
\tsc{WGM}
\tsc{QE}
\tsc{EP}
\tsc{PMS}
\tsc{BEC}
\tsc{DE}

\begin{document}

\shorttitle{SHARP(ening the) view of unresolved globular
clusters in ultra diffuse galaxies}    

\shortauthors{M. Mirabile et al.} 
\title[mode=title]{SHARP(ening the) view of unresolved globular
clusters in ultra diffuse galaxies}


\author[inst1,inst2,inst3]{Marco Mirabile\orcidlink{0009-0007-6055-3933}}[orcid=0009-0007-6055-3933]

\cormark[1]
\ead{marco.mirabile@inaf.it}
\ead[URL]{www.marcomirabile.com}

\author[inst1]{Michele Cantiello\orcidlink{0000-0003-2072-384X}}
\author[inst1]{Benedetta Di Francesco}
\author[inst1]{Ivan Di Antonio}
\author[inst1]{Gianluca Di Rico}
\author[inst5]{Chiara Buttitta\orcidlink{0000-0001-9787-3067}}

\author[inst8]{Marco Gullieuszik\orcidlink{0000-0002-7296-9780}}

 \author[inst9,inst10,inst11]{Johanna Hartke\orcidlink{0000-0002-8745-689X}}
\author[inst5]{Enrichetta Iodice}
\author[inst5,inst6]{Maurizio Paolillo}

\affiliation[inst1]{
  organization={INAF Osservatorio Astronomico d’Abruzzo},
  addressline={Via Maggini}, 
  city={Teramo},
  postcode={64100},
  country={Italy}
}

\affiliation[inst2]{
  organization={European Southern Observatory},
  addressline={Karl-Schwarzschild-Strasse 2},
  city={Garching bei München},
  postcode={85748},
  country={Germany}
}

\affiliation[inst3]{
  organization={Gran Sasso Science Institute},
  addressline={Viale Francesco Crispi 7},
  city={L'Aquila},
  postcode={67100},
  country={Italy}
}
\affiliation[inst5]{
   organization={INAF -- Astronomical Observatory of Capodimonte},   addressline={Salita Moiariello 16},
   city={Naples},
   postcode={80131},
   country={Italy}
 }

\affiliation[inst6]{
  organization={University of Naples “Federico II”},
  addressline={C.U. Monte Sant’Angelo, Via Cinthia},
  city={Naples},
  postcode={80126},
  country={Italy}
}

\affiliation[inst8]{
  organization={INAF - Osservatorio Astronomico di Padova},
   addressline={Vicolo dell'Osservatorio 5},
  city={Padova},
   postcode={35122},
   country={Italy}
 }

\affiliation[inst9]{
   organization={Finnish Centre for Astronomy with ESO (FINCA)},
   addressline={University of Turku},
   city={Turku},
   postcode={FI-20014},
   country={Finland}
}

\affiliation[inst10]{
   organization={Tuorla Observatory, Department of Physics and Astronomy},
   addressline={University of Turku},
   city={Turku},
   postcode={FI-20014},
   country={Finland}
 }

\affiliation[inst11]{
   organization={Turku Collegium for Science, Medicine and Technology (TCSMT)},
   addressline={University of Turku},
   city={Turku},
   postcode={FI-20014},
   country={Finland}
 }


\cortext[cor1]{Principal corresponding author}

\begin{abstract}
Ultra-Diffuse Galaxies (UDGs) challenge current galaxy formation models within the $\Lambda$CDM framework, particularly regarding their dark matter content and formation histories. Globular Clusters (GCs) serve as excellent tracers for probing the halos of these faint systems; however, current spectroscopic facilities struggle to reach the depth required to fully characterize GC populations beyond the local neighborhood. In this work, we present the scientific potential of NEXUS@SHARP, a near-infrared multi-slit spectrograph proposed on the upcoming Extremely Large Telescope (ELT). We demonstrate that NEXUS is uniquely positioned to overcome current observational limitations by combining high spatial resolution with efficient multiplexing. We show that NEXUS can efficiently help to characterize the GC population in UDGs across diverse environments-- out to 60 to 70 Mpc--with a few hours of ELT time. These capabilities will allow for precise measurements of radial velocities and stellar populations, enabling robust dynamical mass estimates and distinguishing between competing UDG formation scenario.
\end{abstract}

\begin{keywords}
ELT\sep
near-infrared spectrograph\sep
Star cluster\sep
Galaxy Cluster\sep
Galaxy Groups\sep
Dwarfs\sep
galaxy:formations\sep
galaxy: evolution
\end{keywords}

\maketitle

\section{Scientific Rationale}
\label{sec:introduction}

The efficiency with which galaxies convert their gas reservoirs into stars depends on their halo mass \citep{Moster13, Behroozi13}. In the absence of astrophysical processes, dark matter (DM) halos of all masses would be expected to contain a baryonic component consistent with the cosmological baryon fraction \citep{Planck20}, $\Omega_{b}/\Omega_{m}  \sim 0.17$, meaning that roughly 17\% of their total mass should initially be in the form of gas and, eventually, stars. This value provides a reference against which the efficiency of galaxy formation can be measured through observables such as the stellar-to-halo mass relation \citep[SHMR,][]{Kostov18}.

In practice, halos retain and convert only a fraction of these baryons into stars due to a variety of internal and external processes that regulate galaxy growth. Gas accretion, radiative cooling, star formation, and feedback from supernovae and active galactic nuclei all modulate how much of the baryonic material can condense into galaxies. Environmental mechanisms such as ram-pressure stripping and tidal interactions further reduce baryon retention, particularly in low-mass systems. These regulatory processes are essential for reconciling theoretical predictions--such as the galaxy stellar mass and luminosity functions in $\Lambda$CDM--with observations \citep{,Somerville15}. Halos with masses comparable to that of the Milky Way are thought to be the most efficient, converting about 20\% of their baryons into stars. At higher masses, internal and external processes suppress gas cooling, producing lower star-formation efficiencies in massive groups and clusters. Conversely, dwarf-scale halos retain only about 1–5\% of their baryons in stars; this drop is mainly driven by stellar feedback and the halos’ shallow potential wells, which make it easier for gas to escape \citep{Bullock17}

Although secondary factors such as galaxy morphology, star formation history, and environment can modulate these trends, their impact is generally modest. Across cosmic time and diverse environments, halo mass remains the dominant parameter governing the overall efficiency of baryon-to-star conversion \citep{Martizzi20}. This mass-dependent efficiency is a cornerstone of modern galaxy formation models and shapes key scaling relations linking baryonic and DM components. Modern cosmological simulations, including IllustrisTNG \citep{Nelson19}, EAGLE \citep{Schaye15}, and FIRE \citep{Hopkins18}, have provided strong support for this framework, generally reproducing the broad observational trends of the SHMR. However, despite this broad agreement, the models diverge at the low-mass end, where the physical processes governing star formation and feedback are most uncertain. In particular, each simulation relies on different sub-grid prescriptions, often calibrated differently and implemented with varying numerical strategies. As a result, their predictions for the low-mass SHMR, baryon cycling efficiency, and dwarf-galaxy star-formation histories can vary by up to an order of magnitude, highlighting the sensitivity of this regime to unresolved physics and the continuing need for improved constraints from both simulations and observations \citep[see][for a review on cosmological models]{Vogelsberger20}.

The last six decades of research on the massive end of the galaxy mass function have established a bottom-up, hierarchical assembly picture \citep{Blumenthal84}. In this paradigm, small DM halos collapse first and later merge into progressively larger systems, implying that low-mass galaxies are expected to reside in these early, small halos. The advent of wide and deep sky surveys has revolutionized this field by enabling detailed studies of low-mass and low-surface-brightness galaxies across environments. Those observations, together with higher-resolution simulations, have revealed several small-scale challenges to the $\Lambda$CDM paradigm --including the cusp-core, missing-satellites, and too-big-to-fail problems \citep[e.g., ][]{Bullock17}.

\begin{figure}
    \centering
    \includegraphics[width=0.5\textwidth]{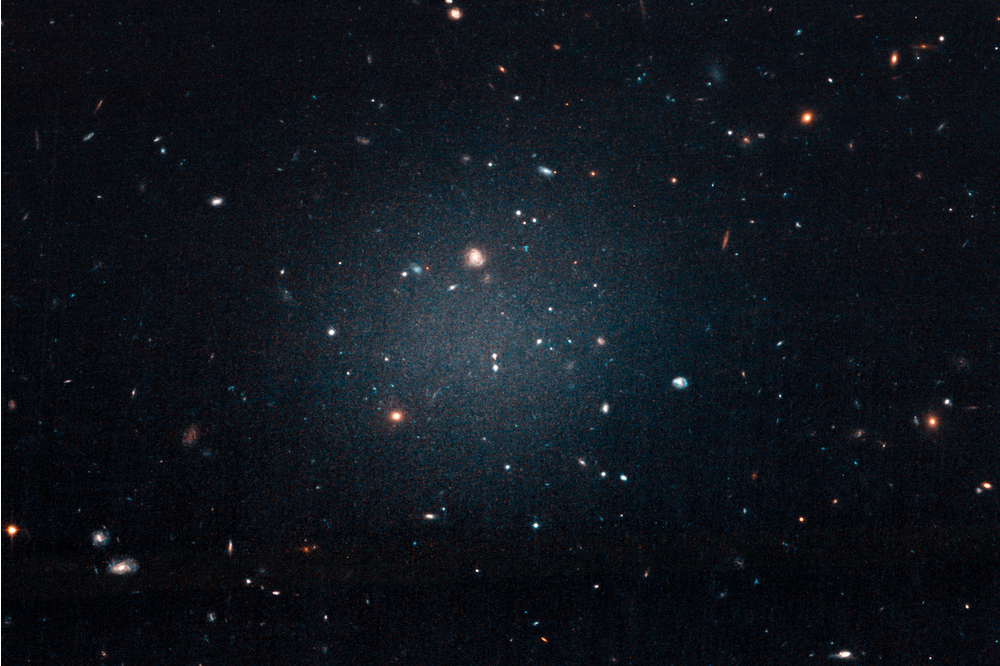}
    \caption{Color composite image obtained with the Hubble Space Telescope of NGC 1052-DF2. Image credit: NASA / ESA / P. van Dokkum, Yale University.}
    \label{fig:UDG}
\end{figure}

Extended low surface brightness galaxies have been known since the 1980s \citep{Sandage84}, but it was only in 2015 that they become popular, when \citet{vandokku15} identified dozens of ultra-diffuse galaxies (UDGs, see Fig. \ref{fig:UDG}) in the Coma cluster. These were defined as galaxies with central surface brightness $\mu_{g,0} \gtrsim 24.0,\mathrm{mag/arcsec^2}$ and effective radii $R_e \gtrsim 1.5,\mathrm{kpc}$--basically galaxies with the size of the Milky Way but about 100 times less massive. Initially thought to predominantly inhabit dense environments \citep{Koda15, vandokku15, Venhola17, Venhola22, lamarca22b, Iodice23, Marleau24a}, UDGs have since been identified in a wide range of environments, including the field \citep{Martinez16, habas20, Muller21, Marleau21, Mirabile24}.

The origin of UDGs remains an open question, with two main formation scenarios. One proposed scenario is that UDGs are "failed" massive galaxies, $L^\ast$ systems quenched early ($z \sim 2$), which preserved dwarf-like stellar masses within relatively massive halos. These galaxies would thus appear red, old, metal-poor, DM dominated, and host unusually rich globular cluster (GC) population \citep{vandokku15, vandoku16, Janssens22, Toloba23}. Alternatively, UDGs may be "puffed-up" dwarfs, genuinely low-mass galaxies that have expanded due to feedback or environmental processes \citep{Amorisco16, DiCintio17, Carleton19, Benavides23, Zheng25}. Relative to the previous scenario, such objects are expected to be bluer, exhibit extended star formation histories, and have DM content and GC population, resembling those of typical dwarfs.

There are also other ways to form UDGs such as external mechanisms including galaxy interactions, mergers, and environmental stripping \citep[e.g.,][]{Ivleva24} Possible origins include collisional debris from galaxy mergers \citep{Lelli2015}, like tidal dwarf galaxies, or gas clumps stripped by ram pressure from infalling systems \citep{Yozin15,Poggianti19,Junais22, Ivleva24}. These scenarios yield blue, dusty, star-forming UDGs with low metallicities, little or no DM, and significant UV emission.

While external mechanisms provide a valuable explanation for galaxies with a small amount of DM, they are unable to account for systems such as NGC\,1052-DF2 \citep{vanDokkum18} and NGC\,1052-DF4 \citep{vandoku19}. These two UDGs appear to exhibit a lack of DM in their inner regions, challenging our understanding of galaxy formation within the $\Lambda$CDM framework \citep{Haslbauer19}. Several explanations have been proposed, with the leading model suggesting that high-velocity interactions separated baryonic and DM while generating the pressure needed to form bright GCs \citep[bullet dwarf scenario,][]{Silk19, vanDokkum22}. These galaxies indeed host unusually bright GCs, with top-heavy GC luminosity functions (GCLFs) that differentiate them from normal dwarfs \citep{vandoku19, Shen21}. \citet{Buzzo25} identified another DM-deficient system, FCC\,24, which shares several traits with NGC\,1052-DF2 and NGC\,1052-DF4, including slow and prolate rotation, quiescence in low-density environments, coeval formation of stars and GCs, flat stellar population gradients, top-heavy GCLFs, and nearly monochromatic GC populations. These shared features suggest the existence and a way to identify a class of DM deficient dwarf galaxies.


In summary, current observations indicate that UDGs likely represent a heterogeneous population encompassing galaxies with distinct intrinsic properties, such as colors, stellar populations, and DM fractions.  

Hence, to understand and discriminate among the different formation scenarios, we should collect information on the stellar population (field stars and GCs) and DM content in these galaxies. An example of a coherent analysis of both the stellar populations and dark‑matter content of UDGs in the same environment is provided by the LEWIS survey, which studied UDGs in the Hydra I cluster \citep[see e.g.,][]{Iodice23,Hartke25,buttitta25,Mirabile25,Doll26,Rossi26}


Here, we focus on the GC population because, among the properties discussed in the next section, GCs act as fossils that trace the history of a galaxy: they serve as tracers of halo mass and of the system's dynamics, providing a direct diagnostic to distinguish among UDG formation channels \citep{Toloba18,Buzzo25b}

\subsection{Globular cluster properties}

GCs are ancient ($t \gtrsim 8$ Gyr), compact stellar systems with typical half-light radii of $r_e \sim 2.5$ pc, formed during the early phases of galaxy assembly when intense star formation favored the creation of massive clusters \citep{Larsen00, Brodie06, forbes2021}. As relics of this epoch, they act as fossil records of the host galaxy formation and evolution. 
The GC systems in external galaxies typically exhibit a consistent luminosity function, which is well approximated by a Gaussian profile with a characteristic peak--known as the turnover magnitude (TOM)--at an absolute magnitude of $M^{\mathrm{TOM}}_V \sim -7.5$ mag \citep{harris01, rejkuba12}. This near-universality allows the GC luminosity function (GCLF) to serve as a reliable secondary distance indicator \citep{ferrarese20}. The width of the GCLF is observed to depend on the host galaxy mass, with more massive galaxies showing broader GCLF \citep{villegas10}. 
The TOM also shows variations across galaxy types. In dwarf galaxies, the TOM is observed to be slightly fainter compared to more massive systems, indicating a dependence on host galaxy mass \citep{villegas10}. The situation is more debated for UDGs. Some studies report a TOM consistent with that of typical dwarf galaxies \citep{Saifollahi22, Li25}, while others find significant offsets \citep{vandoku19,Muller21}. Thanks to their high luminosity, GCs are detectable at cosmological distances \citep[e.g.,][]{alamo13}. The number of GCs hosted by galaxies varies depending on the galaxy type, from dwarf galaxies hosting zero to a few dozen, to spiral galaxies with hundreds, and up to elliptical galaxies where this number can reach several thousand. In the case of UDGs, the number of GCs ranges from zero to even $\sim$100 \citep[e.g., DF-44,][]{vandokkum17}, which is an unexpectedly high number of GCs considering their stellar mass. 

Studying the GC population of UDGs, especially estimating the number of GCs in such systems,  serves as a valuable diagnostic tool to investigate the formation pathways of the UDGs (see Sect. \ref{sec:introduction}). 
Deep imaging campaigns targeting UDGs in clusters and isolated environments \citep{Prole19, Lim18, Jones23} have quantified GC populations, revealing that some UDGs host exceptionally rich GC systems for their stellar mass, suggesting the presence of massive DM halos, while others contain GC populations consistent with ordinary dwarf galaxies \citep{Beasley16, Amorisco18, Lim20, Saifollahi21b}. The total number of GCs in a galaxy correlates with its halo mass \citep{blake97, Harris17, Burkert20}, and is thus providing an indirect tracer of its DM content. This correlation arises because GC formation and survival scale with the depth and assembly history of the host halo, so richer GC systems trace more massive DM halos \citep{Kruijssen26}. At the same time, knowing the total number of GCs (\(N_{\rm GC}\)) allows us to estimate the specific frequency, defined as \(S_N = N_{\rm GC}\,10^{0.4(M_V+15)}\), where \(M_V\) is the galaxy's total V-band magnitude \citep{harris91}. This quantity measures the galaxy’s efficiency in forming and retaining GCs over cosmic time. A scatter in \(S_N\) at fixed stellar mass highlights different evolutionary pathways.



Spectroscopic studies of GCs provide valuable constraints on the evolutionary pathways of UDGs by measuring kinematic signatures that can indicate the influence of tidal interactions and environmental effects \citep{Toloba18}. The velocity dispersion derived from the motions of GCs enables direct estimates of DM content, as higher dispersions generally point to more massive halos \citep{Toloba23, Buzzo25, Haacke25}. However, caution is needed in interpreting these measurements: contamination from unbound or intracluster GCs and the use of small samples can lead to uncertainties and potentially overestimate halo masses, especially for UDGs with fewer than 10 GC candidates \citep{Toloba23, Doppel24}. Nevertheless, in systems where stellar kinematic tracers are absent (mainly because of the prohibitive exposure times eventually required to get spectra for the low surface brightness host galaxy), GCs remain an essential tool for probing the DM halos and testing theoretical predictions for their formation and mass assembly.

\subsection{Open questions and limits of the current observations}

\label{sec:open_que}

\noindent
\textbf{Is the scaling relation between the $N_{GC}$ and halo mass valid for UDGs?}

The validity of the $N_{GC}$–halo mass relation for UDGs remains a subject of active investigation and debate. Although some studies report that the relation holds reasonably well for UDGs hosting rich GC systems \citep[$N_{GC}\geq20$,][]{Forbes25}, notable outliers appear --especially among systems with few GCs. In this low-$N_{GC}$
regime, stochastic sampling, and variations in globular-cluster formation efficiency can produce substantial scatter \citep{georgiev10,harris13,Forbes18}. The diversity of UDG formation channels and environmental histories may further complicate the straightforward application of this scaling relation. 

\noindent
\textbf{What does the $S_N$ of these galaxies tell us? }

\begin{figure}
    \centering
    \includegraphics[width=0.45\textwidth]{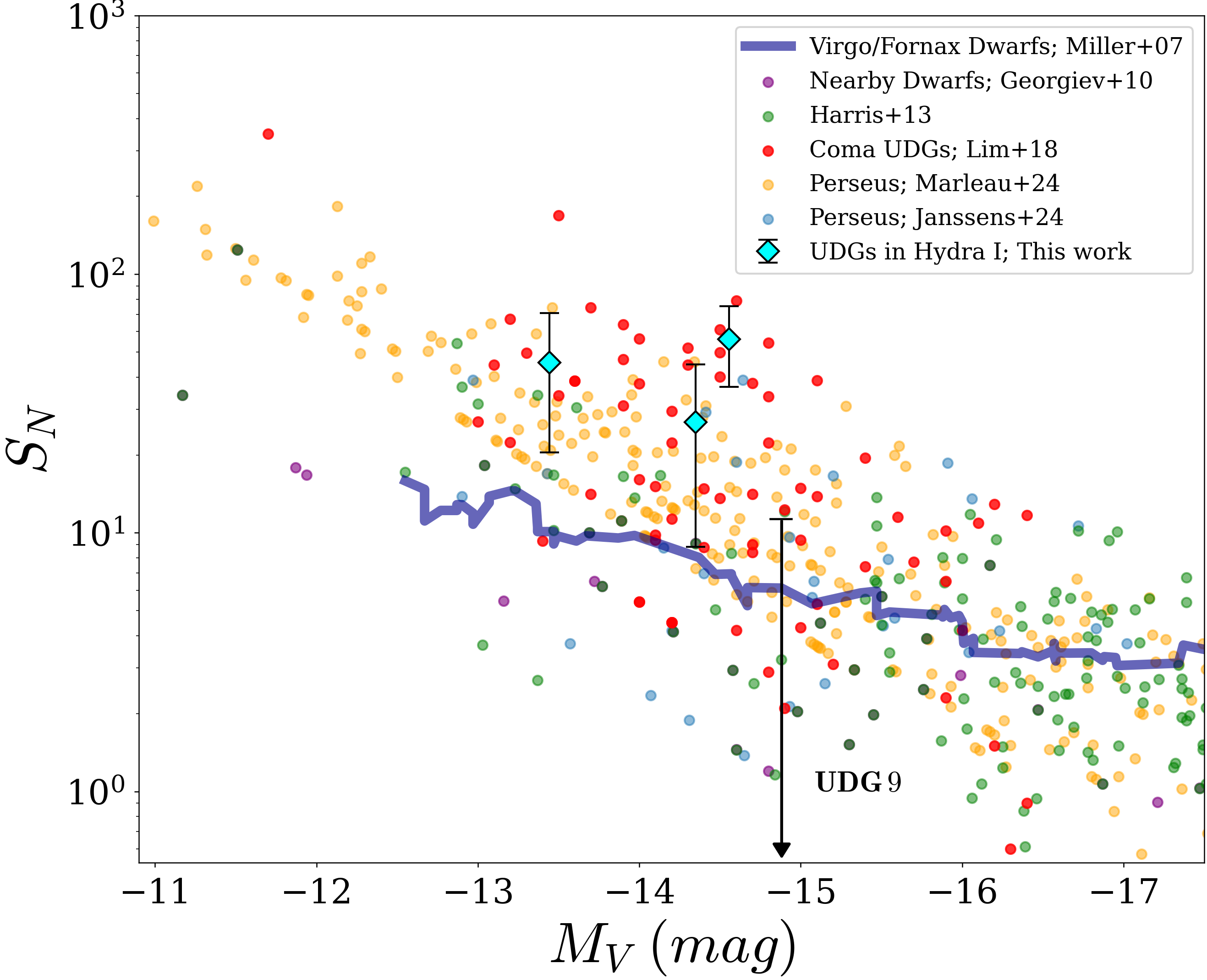}
    \caption{Specific frequency versus magnitude plot comparing Hydra I UDGs $S_N$ (light-blue diamond) with literature values of dwarfs and UDGs from different environments \citep{miller07,georgiev09,harris13,Lim18,Marleau24a,Janssens24}. All literature measurements are shown as colored dots, except for the \citet{miller07} data, which is displayed as a running mean (solid blue line). The black solid arrow represents the position of UDG\,9, a galaxy in the Hydra I cluster for which the population of GCs is statistically consistent with zero. \textit{Adapted from:} \citet{Mirabile25}.}
    \label{fig:SN}
\end{figure}

Figure \ref{fig:SN}, adopted from \citet{Mirabile25}, shows the $S_N$ for different dwarf and UDGs in the literature obtained from mainly photometric studies. At fixed luminosity (mass), $S_N$ shows a significant scatter that may reflect underlying variations in GC formation efficiency, accretion history, and environmental dependencies. However, at low GC numbers, the GC formation process itself becomes stochastic  \citep{georgiev10,harris13}. Additionally, the estimate of the total GC population from imaging is affected by several factors. These include the selection criteria adopted for GC candidates, the statistical background subtraction techniques used to account for contamination from Milky Way stars and background galaxies \citep{cantiello15,dabrusco16,Cantiello18,hazra24,Mirabile24,Lonare24}, as well as observational limitations such as image depth and quality. These effects make direct comparisons across studies more difficult and introduce large uncertainties in the estimated total $N_{GC}$.
Thus, whether the observed scatter at fixed stellar mass is real or the consequence of observational bias remains a matter of debate.

\noindent
\textbf{What is the shape of the GCLF in UDGs?}


Studies of the GCLF in UDGs reveal intriguing diversities, supporting the existence of multiple formation pathways. Several UDGs exhibit dwarf-like GCLFs, with turnover magnitudes consistent with those of typical dwarf galaxies \citep{Saifollahi22,Li25}. In contrast, other systems, such as NGC\,5846-UDG\,1 and VCC\,615, show more luminous, massive galaxy-like GCLFs \citep{Muller21,Danieli22,Mihos25}. Moreover, some dark matter-deficient UDGs, including NGC\,1052-DF2, NGC\,1052-DF4, and FCC\,224, display bimodal GCLFs with an additional bright component associated with UCD\footnote{Ultra compact dwarf (UCD).}-like objects that are significantly more luminous than canonical GCs \citep{vandoku19,Tang25}. This diversity suggests that GC-rich UDGs with luminous or top-heavy GCLFs may have formed in high-pressure environments, consistent with the failed galaxy or bullet dwarf scenarios, while systems with dwarf-like GCLFs are more compatible with standard dwarf galaxy evolution.

However, our understanding of the GCLF in UDGs remains incomplete, especially for the small number of large and deep studies, since many UDGs are faint and distant, making accurate measurement of the GCLF turnover is challenging. Photometric uncertainties, contamination, and limited number statistics complicate the determination of the GCLF parameters. It remains unclear whether the observed variations in the GCLF reflect genuine differences in GC formation or are due to environmental effects, selection biases, or stochasticity.




\noindent
\textbf{Existing spectroscopic campaigns of GC systems in UDGs.
}
Over the last decade, deep wide-field surveys have uncovered large numbers of UDGs \citep[e.g.,][]{Zaritsky19,Marleau21,lamarca22b,Marleau24a,Saifollahi24}, revealing a heterogeneous population with diverse colors, GC richness/populations, stellar populations, and DM fractions. However, as already highlighted, imaging surveys and studies--especially those focused on GCs in such faint systems--ace several challenges. These include dataset incompleteness (e.g., insufficient depth to fully sample the GCLF), selection biases, and poor statistical removal of contaminants to extract reliable measurements. All of these factors can influence the interpretation of results when comparing different studies and may bias our understanding of this intriguing class of galaxies.
One way to produce more coherent and homogeneous results is to obtain spectroscopic information for the GC candidates in these galaxies. Unfortunately, spectroscopy results to be extremely telescope-time-consuming over such the faint extra-galactic GCs, resulting in only a limited number of UDGs having been studied spectroscopically \citep[][and citation therein]{Gannon24}, and even fewer with integral-field (IF) spectroscopy \citep[e.g.,][]{muller20}. For instance, in galaxy clusters, the survey Looking into the faintEst WIth MUSE  \citep[LEWIS,][]{Iodice23,Hartke25,buttitta25,Mirabile25} provided the first homogeneous IF study of 18 UDGs and 6 low-surface brightness (LSB) galaxies in Hydra I galaxy cluster. At the distance of Hydra I ($d\sim50$ Mpc), the LEWIS survey managed to get reliable spectroscopy for the brightest GC component in the observed dwarfs. \citet{Mirabile25} obtained the first detailed study of the GCs over four galaxies in the survey--soon to be complemented with the remaining 20 targets--and identified 11 spectroscopic GCs, either host in the galaxy, or intra-cluster ones. In Virgo ($d\sim16.5$ Mpc), \citet{Toloba23} studied a sample of 10 UDGs and, using GC velocity dispersions, for the first time, identified UDGs with apparently overmassive dark‑matter halos. In Coma ($d\sim100$ Mpc), the spectroscopic analysis of 10 UDGs by \citet{Gannon24} shows that six are GC‑rich, while the remaining four lack GC information.
In the literature, there are other spectroscopic studies targeting UDGs, but only few have acquired spectroscopic information for their GC populations \citep[e.g., ][]{FerreMateu23,Fahrion25} Most spectroscopic works to date have focused on the integrated stellar light of UDGs, often limited to small, brighter samples due to the observational challenges posed by their low surface brightness. As a result, detailed dynamical and chemical characterization of GC systems within UDGs remains largely unexplored, restricting our ability to fully understand their DM 
content and formation histories from these crucial tracers.
This situation underscores the importance of new instruments that will enable simultaneous spectroscopy of multiple GCs within UDGs and their field of stars, expanding the kinematic and population studies in these faint systems.

\section{Objectives}

\begin{table*}
\centering
\caption{Properties of different environments and expected UDG population}
\vspace{2pt}
\begin{tabular}{lccccccc}
\hline
\hline
Environment &$M_{200} (M_\odot)$ & $N_{UDGs}$  & D (Mpc) & $m_V^{\rm TOM}$ (mag)  & MUSE Exp. Time (h)& SHARP Exp. Time (h) \\
 &   (1) &(2) &(3) &(4) &(5) &(6)\\ \hline
Virgo       &$1.05\times10^{14}$&  $\sim20$   &$\sim16.5$& 23.5  &  $\sim7$& $\sim0.08$ (5 min)     \\
Fornax      & $7.0\times10^{13}$& $\sim15$    & $\sim20$ & 24    &  $>100$  & $\sim0.1$ (8 min)  \\
Centaurus/Hydra I     &$2.1\times10^{14}$ & $\sim45$     & $\sim50$ & 26    &  $\gg500$& $\sim5$   \\
Perseus*     & $6.65\times10^{14}$& $\sim150$    & $\sim70$ & 26.8       & $\gg10^3$ & $\sim20$    \\
\hline
\hline
\end{tabular}

\begin{justify}
    Notes:  Col. 1 Galaxy cluster $M_{200}$ mass taken respectively from \citet{Simionescu17,drinkwater01,Tamura00,Simionescu11}; Col. 2 is the number of UDGs ($\propto M_{200}^{1.11}$) expected in each cluster using Eq. 1 from \citet{vanderBurg17}; Col. 3 lists the distance to each cluster: values for Virgo and Fornax are taken from \citet{blakeslee09}, for Hydra from \citet{Christlein03}, and for Perseus from \citet{tully09}; Col. 4 is the expected GCLF TOM; Col. 5 reports the expected exposure time to reach the TOM with MUSE assuming a G2V Pickles model, with a spatial coadding of $4\times4$ pixels, a spectra coadding of 2 pixels, with airmass $sec(z)\leq1.5$ and a seeing $\leq0.85\ \mathrm{arcsec}$; Col.6 reports the expected exposure time to reach the TOM using NEXUS/SHARP@ELT.
    *Not observable from Cerro Armazones; here reported only for reference.
    \end{justify}
\label{tab:ETC}
\end{table*}

To shed light on the nature of UDGs, deep spectroscopic observations of a large sample of UDGs are needed, as combining the analysis of GC richness with the kinematic and evolutionary properties of field stars (age and [Fe/H])  provides a powerful diagnostic for distinguishing between different formation channels. Such observations also offer the opportunity to measure the DM content of these galaxies independently from their stellar kinematics, GC counts, and  GC kinematics \citep[e.g.,][]{Toloba23, buttitta25, Mirabile25}. Here, we focus on the insights that can be gained from studying the GC population.

\subsection{Perspectives with SHARP}
\label{sec:sharp}

SHARP is a near-infrared ($0.95-2.45\ \mu m$) spectrograph \citep{Saracco25,Mahmoodzadeh25} specifically designed to leverage the capabilities of MCAO\footnote{MCAO stands for multi-conjugate adaptive optics.}-assisted observations at the  ESO's Extremely Large Telescope (ELT), through the MORFEO adaptive optics system \citep{Ciliegi21}. The instrument comprises two complementary units: NEXUS, a multi-object Spectrograph, and VESPER, a multi-Integral Field Unit system. With a pixel scale of 35 mas (NEXUS) and 31 mas (VESPER), SHARP achieves spatial sampling of $\sim$150-250 pc (the scale of giant molecular clouds) across cosmic time, surpassing JWST's angular resolution by factors of 3-6. The instrument characteristics are described in more detail by Saracco et al. in this volume. In Sect. \ref{sec:spec_gc} we give the specific SHARP configuration recommended for the science case discussed here.

To address the number of open questions on UDG formation (see Sect. \ref{sec:open_que}), particularly through their GC component, deep spectroscopic observations sampling the GCLF at least down to the TOM for a large number of UDGs are necessary. Such observations could provide robust measurements of the total $N_{GC}$
in each galaxy by confirming cluster membership through radial velocity estimates. In Table \ref{tab:ETC}, we report a list of prototypical galaxy clusters at increasing distance that, if targeted by deep spectroscopic observations with SHARP, could allow to study a large number of UDGs due to their high richness. Nevertheless, isolated UDGs remain of great interest for constraining their possible formation channels.

A key motivation for reaching at least the GCLF turnover magnitude is that doing so enables a complete and unbiased census of the GC population in each UDG. Accurate determination of $N_{\mathrm{GC}}$ is essential for testing the empirical scaling relation between GC number and halo mass. 
Additionally, it allows us to determine whether UDGs conform to the “universal” GCLF shape, or whether they display systematic differences--such as a deficit of low-mass clusters, excess of luminous clusters, or a shift in the turnover luminosity. 
Moreover, the stellar populations of the GCs can be investigated, providing crucial information on their ages and metallicities. This allows to connect the dynamical and luminosity properties of the GCs with their formation history, helping to disentangle whether observed variations in the GCLF are driven by intrinsic differences in GC formation, subsequent evolution, or environmental effects within the host UDGs. So, the combination of accurate $N_{\mathrm{GC}}$ measurements,  dynamical mass constraints, and the GCLF properties provides a powerful multi-dimensional diagnositc of the evolutionary pathways of UDGs through their GCs, helping to determine whether they represent faded dwarfs, failed massive galaxies, environmentally altered systems, or a mixture of these scenarios.


Deep spectroscopic data of GCs in UDGs will not only deepen our understanding of GC systems in currently known UDGs, but will also exploit and complement previous observational efforts. For example, deep imaging surveys \citep[e.g.,][]{ferrarese12} with HST, Subaru, and CFHT have provided initial catalogs of GC candidates, but without spectroscopic confirmation, these studies are limited by contamination, distance uncertainties, and incomplete sampling of the GCLF. At the same time, deep spectroscopy will also create strong synergies with ongoing and upcoming facilities and surveys such as Euclid, LSST, and JWST. These missions are dramatically increasing the number of UDGs and other low-surface-brightness systems accessible for study, over wide areas and to unprecedented depths, enabling the construction of catalogs of GC-rich low-surface-brightness galaxies for targeted follow-up with deep spectroscopic observations.

However, obtaining such detailed spectroscopic data for faint systems at these distances presents a significant challenge for current facilities. In the following section, we describe the observational setup and the specific instrument capabilities required to achieve these objectives efficiently, with particular attention to the methods and possibilities allowed by a unique facility like SHARP.

\subsection{Spectroscopy of GCs in UDGs: Existing observing facilities}
\label{sec:spec_gc}

With the aim of observing GC populations in UDGs, we present here the case study to target multiple GCs simultaneously to capture GCLF, measure radial velocities, and characterize their stellar populations. Among the best available ground-based instruments for this purpose, we considered MUSE on the ESO VLT, which can target multiple GCs simultaneously and cover the spectral features needed to measure radial velocities and stellar population properties.

\begin{figure}
    \centering
    \includegraphics[width=0.5\textwidth]{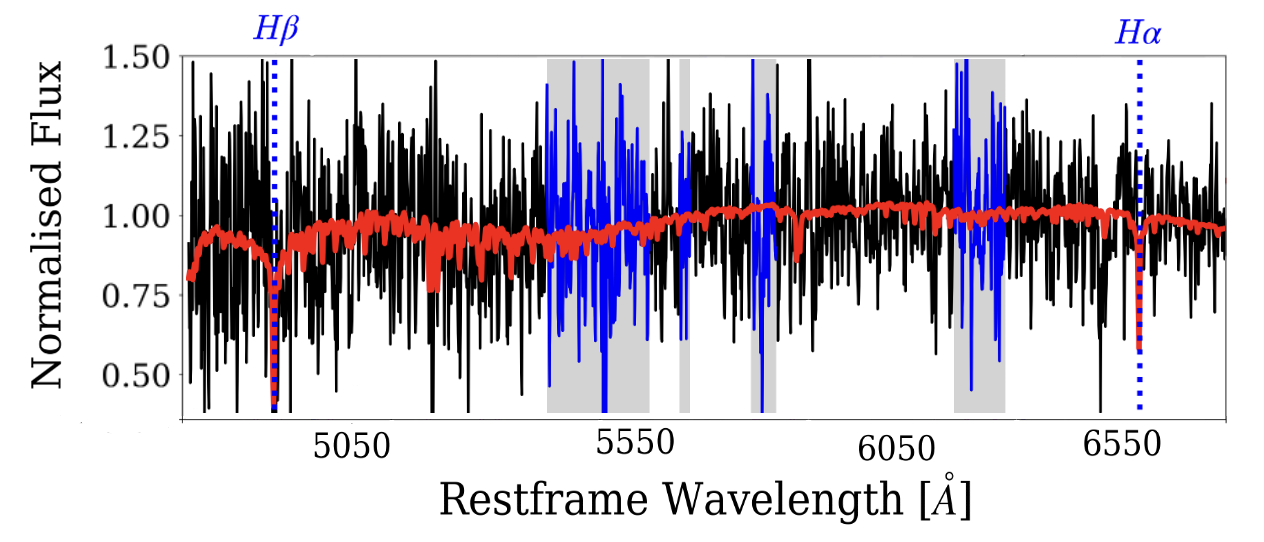}
    \caption{Spectrum of a GC (black line) with SNR of $7\ \mathrm{\mathring{A}^{-1}}$, along with the pPXF best-fit model (red solid line). Regions with strong sky residual lines are masked from the fit (grey shaded areas). The main absorption features ($\mathrm{H_{\alpha}}$ and $\mathrm{H_{\beta}}$) are drawn with blue vertical dotted lines. Figure and caption adapted from: \citet{Mirabile25}.}
    \label{fig:GC}
\end{figure}


The integration times required to reach the depth needed to sample the GCLF down to the TOM at distances typical of nearby clusters are listed in Table~\ref{tab:ETC}. \citet{Mirabile25} showed that even \(\sim6\) hours of exposure with MUSE are insufficient both to reach the GCLF turnover magnitude for UDGs at the distance of the Hydra~I cluster (\(\sim50\) Mpc; \citealt{Christlein03}) and to obtain spectra with adequate S/N for robust analysis, highlighting the difficulty of building statistically complete GC samples with current-generation facilities. Figure~\ref{fig:GC} presents an example GC spectrum observed with MUSE@VLT in Hydra~I for a source of \(m_H\simeq22.5\)~mag \citep{Mirabile25}; the red curve shows the pPXF fit \citep{Cappellari17}. The figure clearly illustrates the low S/N achieved for such sources and underscores the observational difficulty of using advanced IFUs such as MUSE at the VLT, in spectroscopic studies of GC populations at these distances.

On the other hand, future ELT instruments such as MICADO \citep{Davies18}, HARMONI \citep{Thatte24}, and ANDES (see \url{https://andes.inaf.it/}) will offer powerful spectroscopic capabilities but limited multiplexing. MICADO will offers long‑slit coverage, HARMONI is a single‑IFU spectrograph, with a FOV that reaches a few square arcseconds across, and ANDES will provide a high‑resolution single long slit plus IFU mode. These modes either provide only single‑slit coverage or IFU fields of only a few arcsec\(^2\), so they cannot simultaneously capture multiple GCs across the typical angular extent of a UDG; this limits their efficiency for building statistically complete GC samples.

An exception is MOSAIC@ELT \citep{Pello24}, a fiber‑fed multi‑object spectrograph, designed to obtain spectra for up to 200 targets within a FOV \(\sim 40\ \mathrm{arcmin}^2\). However, MOSAIC will operate with a Ground Layer Adaptive Optics system, which corrects only for the ground layer of atmospheric turbulence and therefore delivers a more modest improvement in image quality compared to the full MCAO correction provided by MORFEO for SHARP. Similarly, next‑generation telescopes such as the Giant Magellan Telescope (GMT) \citep{Burgett24} and the Thirty Meter Telescope (TMT), will host optical/near-infrared multi‑object spectrographs with larger fields of view (several arcminutes), which will be suitable for the study of GCs in UDGs. However, because GMT and TMT have smaller primary mirrors than the ELT, reaching the same S/N on the faintest GC targets will require longer exposure times.

Within this landscape, the NEXUS@SHARP/ELT concept occupies a valuable niche. Briefly, to highlight the main characteristics of interest for the scientific case proposed here, NEXUS is a near-infrared multi-slit spectrograph with adaptive optics (AO) correction, offering a field of view of approximately 1.2$'\times$1.2$'$ and a fine pixel scale of $\sim$0.035 arcsec/pixel. It covers a broad wavelength range from 0.95 to 2.45 $\mu m$, divided into four channels roughly corresponding to the Y, J, H, and K bands. Its configurable slit system can deploy $\sim$30 slits (incidentally a good match for the expected GC\footnote{The average expected GC popualtion for GC-rich UDgs is of $\sim20$.} population in GC-rich UDGs) of variable width along two possible axes, with adjustable lengths. NEXUS achieves spectral resolutions of R$\sim$6000, $\sim$2000, or $\sim$300, making it flexible for both kinematic and stellar population studies.

NEXUS results to be ideally suited for GC studies in UDGs because it combines high spatial resolution, thanks to AO correction with efficient multiplexing. Its multi-slit capability allows simultaneous observation of multiple GCs within the typically compact fields of UDGs, substantially improving observational efficiency. The AO correction delivers diffraction-limited performance, enabling precise measurements of faint clusters. 
Together, the capabilities of NEXUS@SHARP and the large collecting area of the ELT enable the study of GC systems in galaxies out to distances of  $\sim$ 60 to 70  Mpc within reasonable observing times.

\subsection{Observational requirements for NEXUS}

Using the SHARP Exposure Time Calculator\footnote{\url{https://sharp.lambrate.inaf.it/}} (version 0.6), we estimated the exposure times required for spectroscopic observations of GCs in UDGs. These calculations are based on stellar population models from \citet{Bruzual03}, assuming a metallicity of Z = 0.004, an age of 10 Gyr, and a Chabrier initial mass function \citep{Chabrier03}. We assumed a spectral resolution R = 2000,  airmass $sec(z)\ \leq$1.5, and a slit width of  $200\ \mathrm{mas}$. The target signal-to-noise ratio, obtained at the reference wavelength of $1.6\ \mu m$,  was set to $\sim$5 at the TOM of the GCLF, which corresponds to detecting half of the GC population.

Last column in Tab. \ref{tab:ETC} summarises the resulting exposure times for NEXUS as a function of distance, providing a practical guide for planning observations that can reliably detect and characterise GC populations in UDGs. These estimates highlight the efficiency of NEXUS/SHARP@ELT to achieve the required depth to reach the TOM in relatively short exposure times compared to current-generation facilities, enabling the sampling of the GCLF up to the TOM, measurement of radial velocities, and characterization of stellar population properties for multiple GCs within a single pointing. 


In summary, simulations with the current NEXUS ETC indicate that NEXUS will enable GC studies in UDGs out to distances at least three times larger than those currently accessible--probing a volume of the local Universe roughly ten times bigger. A single NEXUS pointing will provide radial velocities, total GC counts, and stellar‑population information for multiple GCs, enabling statistically meaningful samples of UDGs across environments (clusters and the field). Measurements of GC system radial velocities and total numbers will permit independent estimates of galaxy dark‑matter content and tests of scaling relations (see Sect. \ref{sec:open_que}). Comparing GC ages and metallicities with those of the host galaxy stellar population can help discriminate between different formation channels: in the failed‑galaxy scenario GCs and field stars should show similar ages and metallicities, whereas in a puffed‑up/diffuse dwarf origin GCs are expected to be systematically more metal-poor than the host stellar population \citep[e.g.,][]{Gannon26}.


\section{Conclusion}

As described in the introductory Section, UDGs can be efficiently investigated through the spectroscopic characterization of their GC populations. Present studies, however, are limited to the nearby Universe within $\sim 20$ Mpc, as larger distances require prohibitive telescope efforts even for a single target. Probing deeper, thus sampling larger volumes of the local Universe and increasing the sample size across a broader range of host environments (clusters, groups, interacting regions, outskirts, field, etc.), is a mandatory requirement for fully constraining galaxy formation models at the faintest surface brightness levels.
In this work, we have demonstrated that the proposed NEXUS@SHARP spectrograph at the Extreme Large Telescope represents a transformative step forward for this field. Our proposed scientific case analysis indicates:

\begin{enumerate}
    \item  Feasibility: NEXUS can reach the GCLF turnover magnitude ($M_V^{TOM} \sim -7.5$) with signal-to-noise ($\sim$ 5) in remarkably short exposure times (from the $\sim$5 minutes at the distance of Virgo, to $\sim$5 hours for Centaurus/Hydra I UDGs), a result unachievable with current facilities.
    \item  Efficiency: The combination of Adaptive Optics and multi-slit capabilities allows for the simultaneous characterization of multiple GCs within a single UDG, drastically reducing the telescope time required for statistical completeness.
    \item Scientific Impact: These observations will provide the necessary kinematic data (e.g. GC system velocity dispersion) and total $N_{GC}$ to derive dynamical masses free from many of the assumptions that affect stellar mass–based estimates. The combination of precise $N_{\mathrm{GC}}$ measurements and dynamical mass determinations therefore allows a self-consistent evaluation of the scaling relation within the same dataset. Agreement between the halo masses inferred from GC counts and from dynamical modeling would reinforce the validity of the scaling relation in UDGs. Conversely, any systematic discrepancies would immediately point to a breakdown of this relation, offering new physical insight into GC formation efficiency, halo assembly, and baryon–DM coupling in low-density galaxy environments. Additionally, these observations will help to understand the nature of the GCLF in these galaxies, and resolve the debate over the UDGs formation scenario.
\end{enumerate}

By enabling the systematic spectroscopic survey of GCs in low-surface-brightness systems, NEXUS@SHARP will act as a definitive tool for constraining the formation channels of the most elusive galaxy populations in the local universe.

\section*{Acknowledgments}
The SHARP team acknowledges support by Bando Ricerca Fondamentale INAF 2022, Techno-Grant "SHARP" - 1.05.12.02.01 and Bando Ricerca Fondamentale INAF 2024, Large-Grant "SHARP" - 1.05.24.01.01.
Part of this work was supported through the INAF “Astrofisica Fondamentale” GO-grant 2024 n. 12 (PI: M. Cantiello). MM acknowledges support form the ESO Studentship Program.
The authors thank \citet{Gannon24} for the compilation of their catalogue of UDG spectroscopic properties. The catalogue includes data from: \citet{mcconnachie12,vandokku15,Beasley16, Martin16, Yagi16,Martinez16,vandoku16,vandokkum17,Karachentsev17,vanDokkum18,Toloba18,Gu18,Lim18,RuizLara18,Alabi18,FerreMateu18,Forbes18,Martin19,Chilingarian19,Fensch19,Danieli19,vandoku19,Torrealba19,iodice20,Collins20,muller20,Gannon20,Lim20,Muller21,Shen21,Ji21,Huang21,Gannon21,Gannon22,mihos22,Danieli22,Villaume22,Webb22,Saifollahi22,Janssens22,Gannon23,FerreMateu23,Toloba23,Shen23}.

\bibliographystyle{A&A_bib_style}
\bibliography{Example}






\end{document}